\documentclass[aps,pre,twocolumn,floatfix,nofootinbib,amsmath,amssymb]{revtex4-2}
\usepackage{times}
\usepackage[sort&compress]{natbib}
\usepackage{graphicx}
\usepackage{amsmath}
\usepackage{graphicx}
\usepackage{subfigure}
\usepackage{xcolor}
\allowdisplaybreaks

\begin{document}

\title {The non-perturbative phenomenon for the Crow Kimura model with  stochastic  resetting}
	\author{R. Poghosyan$^{1,2}$}
	\author{R. Zadourian$^{3}$}
	\author{David B. Saakian$^{2}$}

\email{saakian@yerphi.am }

\affiliation{$^1$
  Shanghai Jiao Tong University,School of Electronic,  Information and Electrical Engineering,
3-317 SEIEE Building, 800 Dong Chuan Rd, Shanghai, 200240, P. R. China }
\affiliation{$^2$
	A.I. Alikhanyan National Science Laboratory
	(Yerevan Physics
	Institute) Foundation,\\
	2 Alikhanian Brothers St., Yerevan 375036, Armenia}

\affiliation{$^3$Wuppertal University, D-42119 Wuppertal, Germany
   }
\date{\today}

\begin{abstract}
We consider the Crow Kimura model, modified via stochastic resetting.
There are two principally different situations:
First, when due to resetting the system jumps to the low fitness state, everything is rather simple in this case, we have a solution {which is} a slight modification
of the standard Crow-Kimura model case.
When there is resetting to the high-fitness state, there is a non-perturbative phenomenon via the resetting probability -- even a minimal 
resetting probability drastically changes the solution.
We found two {subphases} 
in this phase.

\end{abstract}

\maketitle
\medskip

\vskip 3 mm



\maketitle

\section{Introduction}

Stochastic resetting is one of {the} important directions of modern statistical physics \cite{ev20}.
The resetting research started a couple of decades ago while looking at birth-death processes  \cite{br82,br85, ki94}  and nowadays has numerous applications \cite{ev20}
{in fields such as}
ecology  \cite{bo18}, biomedicine   \cite{ra20} and population genetics related  models  \cite{si18},  \cite{fr18},\cite{fr22},\cite{be22}.
As has been observed in the latter works,
the main equation of statistical physics of random resetting processes have been derived already in population genetics literature  \cite{fl79}, while 
{considering} the continuous version of 
 \cite{ki73}.

The idea of resetting search is very simple. 
If we are looking {for} some lost {object},
{then after} starting {the stochastic search} from some initial position,
and {failing} in our search process, 
we renew our search returning back to the starting position.
It is interesting that such strategies are very successful both in nature and in computer algorithms, 
i.e. in case of simulated annealing.

In the standard random walks, one {looks at a} 
linear master equation for the probability distribution, which can be considered as a discrete space 
version of the diffusion {equation}.
There are random walks which are sometimes accompanied with jumps {into} a special state.
In case of {an} absorbing state, the resetting brings to the faster relaxation to the absorbing state, which has many {practical applications}.

{On}
the other hand, we have a problem of nonequilibrium statistical physics, {which is solvable}.
The resetting acts against {reaching} 
equilibrium.

While the equilibrium statistical mechanics already has been successfully applied to the evolution phenomenom {in}
static {conditions}, 
we assume that the nonequilibrium statistical physics {approach}
is a much more adequate language to describe the living matter than the equilibrium statistical physics.
There have been {several}
relating 
stochastic resetting with 
evolution and ecology, working with continuous differential equations.
In resetting problems one works with diffusion or the Fokker Planck equation, which are linear differential equations in 
continuous space.
{On the} contrary, 
in Crow-Kimura \cite{ck70,ba97,sa04} and Eigen \cite{ei71,ei89} {models of evolution}
with an infinitely large population, we work with the nonlinear equations,
{where} 
now we have 
selective forces in addition to 
diffusion, and a discrete set of types.
{ The Crow-Kimura model is related to the discrete time version of the Eigen model, the latter is similar to the the branching processes \cite{sa15}. The Wright Fisher model, which is  the main model of population genetics, is related to the Crow-Kimira model.
So by solving the Crow-Kimura model with resetting, we can extend later our results to the other models of population genetics. The advantage of the Crow-Kimura model is that it is simpler case for the analytical investigation of evolutionary dynamics.}

The vast majority of the stochastic resetting investigations are for 
{Markovian}
models.
Here we introduced the {concept of} resetting for the models with
fitness.
In the case of evolution with a large population, the dynamics is affected by two forces, 
namely selection and mutation.
Without mutations, the population is focused on the state with maximum fitness, 
while a nonzero mutation 
distributes the population around the state with the high fitness.

It is interesting {to study} how the situation changes in our case with resetting, 
similar to \cite{fl79, si18}.
In the latter works, one looks at many alleles in the same locus, approximating the situation 
by the continuous space.
We will formulate the Crow-Kimura model with resetting, and look two different situations 
when resetting is into a state {with} the non-high fitness, 
and the case of resetting to the high-fitness state.
So we work in a discrete space of types, which is more adequate to the biological reality.
We will identify two different statistical physics phases.

\section{The model}
\subsection{The standard Crow-Kimura model}
While for the continuous time random walk probabilities 
we have the diffusion equation for the probability distribution,
\begin{eqnarray}
\label{e} \frac{dP(x,t)}{dt} = D\frac{d^2P(x,t)}{dx^2} \nonumber
\end{eqnarray}
for the resetting case with a resetting rate $\epsilon$ to the position $X_r$,
{the equation is modified into}
\begin{eqnarray}
\label{e1} \frac{dP(x,t)}{dt} =D\frac{d^2P(x,t)}{dx^2}- \epsilon P(x,t)+ \epsilon \delta (x-X_r).
\end{eqnarray}

Consider now the Crow-Kimura model.
We have a genome as a chain {of} letters taking values $\pm$ with length $L$, 
thus we have $2^L$ sequences.
We denote different sequences via an index $0\le i\le 2^L-1$. The Hamming distance $d(i,j)$ between two sequences is the {number of} differences in the signs.

There is a mutation from the $i^{\rm th}$ to the $j^{\rm th}$ state with a rate $\mu_{ij}$.
The latter is nonzero only when $d(i,j)=1$.
We have a mutation rate $\mu/L$ neighbors with the Hamming distance 1, also $\mu_{ii}=-\mu$.

\begin{eqnarray}
\label{e2} \frac{dp(i,t)}{dt} =
(r_i-\mu-R)p(i,t)+\mu\sum_j p(j,t),
\end{eqnarray}
where the sum is over the neighbors with the Hamming distance $d=1$, and $R=\sum_j p(j,t)r_j$.

For the symmetric fitness case, when the fitness is a function of the total number of mutations 
from the $0^{\rm th}$ sequence, and we introduce a fitness function $f(m)$,
$r_j=f(m), m=1-2l/L, l=d(j,0)$. The we define the total probability  $P(l,t)$ of the $l^{\rm th}$
Hamming class:
\begin{eqnarray}
\label{e3}& \frac{dP(l,t)}{dt} = 
(f(m,t)-\mu-R)P(l,t)+\nonumber\\
&\mu(\frac{(L-l+1)}{L}P(l-1,t)+\frac{(l+1)}{L}P(l+1,t)).
\end{eqnarray}
The coefficients $\frac{(L-l+1)}{L}$ and $\frac{(l+1)}{L} $ arose from the combinatorics, 
while considering the Hamming class probabilities.

{ We will investigate the stochastic resetting modification of the model.
While the model is formulated and investigated for the general case of function f(x), in literature three versions of  fitness functions have been popular: a. the single peak case,
where $f(1)=J>0$ while $f(m)=0,m<1$  \cite{ei89}, b. linear fitness function f(x)=kx from population generics \cite{ck70},
c. quadratic fitness function \cite{ba97}.}
\subsection{The stochastic resetting version of the Crow-Kimura model}

Consider the resetting in the Crow-Kimura model to the given configuration $r$.

\begin{eqnarray}
\label{e4} \frac{dp(i,t)}{dt} =
(r_i-\mu-R-\epsilon)p(i,t)+\mu\hat\sum_j p(j,t),
\end{eqnarray}
for $i\ne r$
and for the $r^{\rm th}$ sequence 
\begin{eqnarray}
\label{e5}& \frac{dp(r,t)}{dt} =\nonumber\\
&(r_i-\mu-R-\epsilon)p(r,t)+\mu\hat\sum_j p(j,t)+\epsilon\sum_j p(j,t),
\end{eqnarray}

Consider now the symmetric fitness case, and choose as the resetting state $r$ the $L^{\rm th}$ Hamming class
\begin{eqnarray}
\label{e6}
& \frac{dP(l,t)}{dt}= 
(f(m,t)-\mu-\epsilon)P(l,t)+
\mu(\frac{L-l+1}{L}P(l-1,t)\nonumber\\
&+\frac{l+1}{L}P(l+1,t)), l\ne L \nonumber\\
&\frac{dP(L,t)}{dt} =
(f(-1)-R-1-\epsilon)P(L,t)+
\mu\frac{1}{L}P(L-1,t)+\epsilon, \nonumber\\
&m=1-2l/L
\end{eqnarray}
Thus we have a resetting to the $L^{\rm th}$ Hamming class.

Further we {set} $\mu=1$, which is always possible with the rescaling of the fitness and $\epsilon$.
{ We denote the steady state distribution as $\rho(l)=P(l,T),T>>1$.}

Consider the case when $f(m)\ne f(-m)$ 
and the maximum is at the point $m=1$.
First we obtain $p_L$ in the steady state.
Ignoring $1/L$ terms, we get
\begin{eqnarray}
\label{e7} \rho
(L)=\frac{\epsilon}{R-f(-1)+1+\epsilon}.
\end{eqnarray}
We have done numerics (see Figures 1--3), and identified three different situations.
In the first case, we have a peak at an intermediate Hamming class, 
and the second half-peak near the $L^{\rm th}$ class.
In the second case, we have a single peak near the $L^{\rm th}$ sequence.
In the third case, we have a {half-}peak at the $L^{\rm th}$ sequence.

\begin{figure}[h!]
	\center
	\includegraphics[scale=0.5]{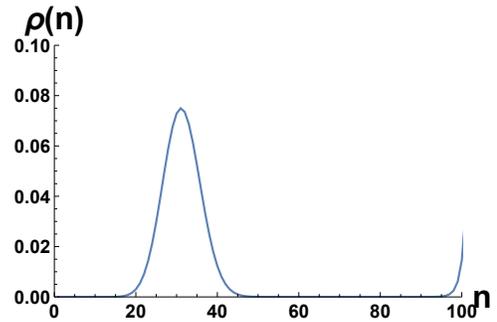}
	\caption{ 
The probability distribution via Hamming classes $\rho(n)$.
 $\epsilon=0.2$, 
  $f(x)=x$, $L=100$.
	}
\end{figure}

\begin{figure}[h!]
	\center
	\includegraphics[scale=0.5]{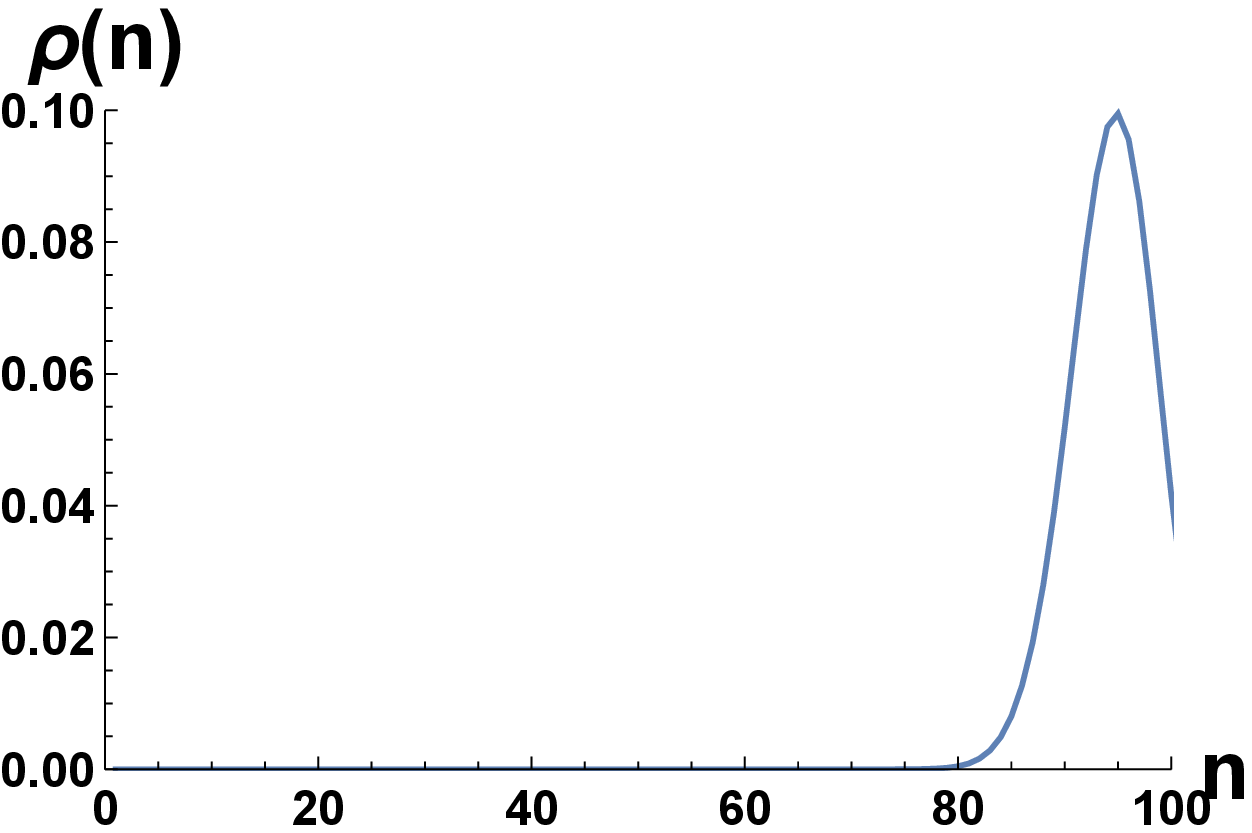}
	\caption{ 
The probability distribution via Hamming classes $\rho(n)$.
 $\epsilon=0.01$, 
  $f(x)=1.5 x^2$, $L=100$.
	}
\end{figure}

\begin{figure}[h!]
	\center
	\includegraphics[scale=0.5]{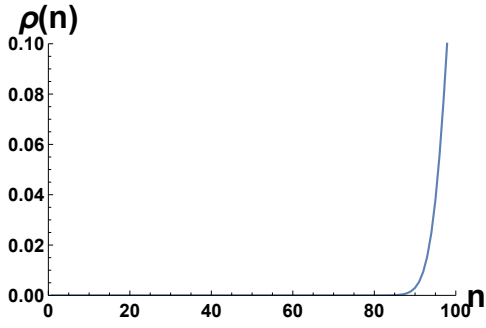}
	\caption{ 
The probability distribution via Hamming classes  $\rho(n)$.
 $\epsilon=0.2$, 
  $f(x)=1.5x^2$, $L=100$.
	}
\end{figure}

\section{The solution of Crow-Kimura model with resetting}

\subsection{The case of resetting to the low fitness state}

Let us express the probability of the $9L-1)$-th sequence via probability of the L-th sequence:
\begin{eqnarray}
\label{e8} 
\rho(L-1)=\frac{  \rho(L)}{R-f(-1)+1+\epsilon}.
\end{eqnarray}
For the Hamming classes $\rho(L-l), l>1$ we get
\begin{eqnarray}
\label{e9} 
\rho(L-l)= \rho(L) \left( \frac{1  }{R-f(-1)+1+\epsilon} \right)^l.
\end{eqnarray}
Summing the members of the geometric progression, 
we obtain for the $P$, the total probability near the $L^{\rm th}$ sequence
\begin{eqnarray}
\label{e1o} P=\frac{\epsilon}{R-f(-1)+\epsilon}.
\end{eqnarray}
There is the second peak of the distribution with the total probability of population around the peak,
{\begin{eqnarray}
\label{e11} 
p=1-P
\end{eqnarray}}
An ansatz \cite{sa07}
\begin{eqnarray}
\label{e12} 
\rho(l)=p \exp( Lu(x))
\end{eqnarray}
gives the Hamilton-Jacobi equation \cite{sa07}
 \begin{eqnarray}
\label{e13}
\frac{\partial u}{\partial t}=f[x]-\epsilon
+\frac{1+x}{2}e^{2u'}+\frac{1-x}{2}e^{-2u'}.
\end{eqnarray}

We obtain the steady state distribution as
   \begin{eqnarray}
\label{e14}
R=f[x]-\epsilon
+\frac{1+x}{2}e^{2u_0'}+\frac{1-x}{2}e^{-2u_0'}-1,
\end{eqnarray}
{ and define define the mean fitness R looking at the maximum of the right hand side via $u'_0$
   \begin{eqnarray}
\label{e15}
R=\max[f[x]-\epsilon+\sqrt{1-x^2}-1
\end{eqnarray}

}

The maximum is at the point $s$ {defined via} 
the equation
   \begin{eqnarray}
\label{e16}
R=f(s)-\epsilon
\end{eqnarray}
while for the standard surplus
   \begin{eqnarray}
\label{e17} 
S=\sum_l \rho(l)(1-2l/N)=ps-(1-p).
\end{eqnarray}
{The surplus is one of the key characteristics of the population distribution, it defines the mean number of mutations as $L(1-S)/2$.}
Our Eqs.(10), (15)-(17) are among the main results of the article.
We compare our analytical results with the numerics.
\begin{figure}[h!]
	\center
	\includegraphics[scale=0.5]{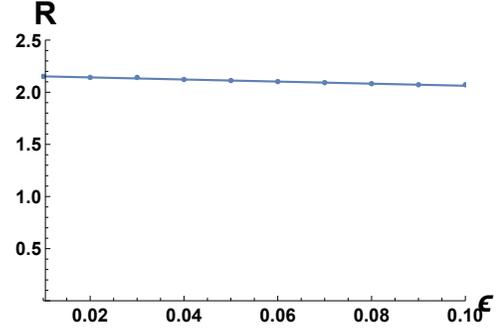}
	\caption{ 
The mean fitness  $R$ versus
 $\epsilon$, 
  $f(x)=3x$, $L=500$.{The smooth line is given by numerics, the solid dots are the theoretical results by Eq. (10),(15)-(17).}	
	}
\end{figure}

\begin{figure}[h!]
	\center
	\includegraphics[scale=0.5]{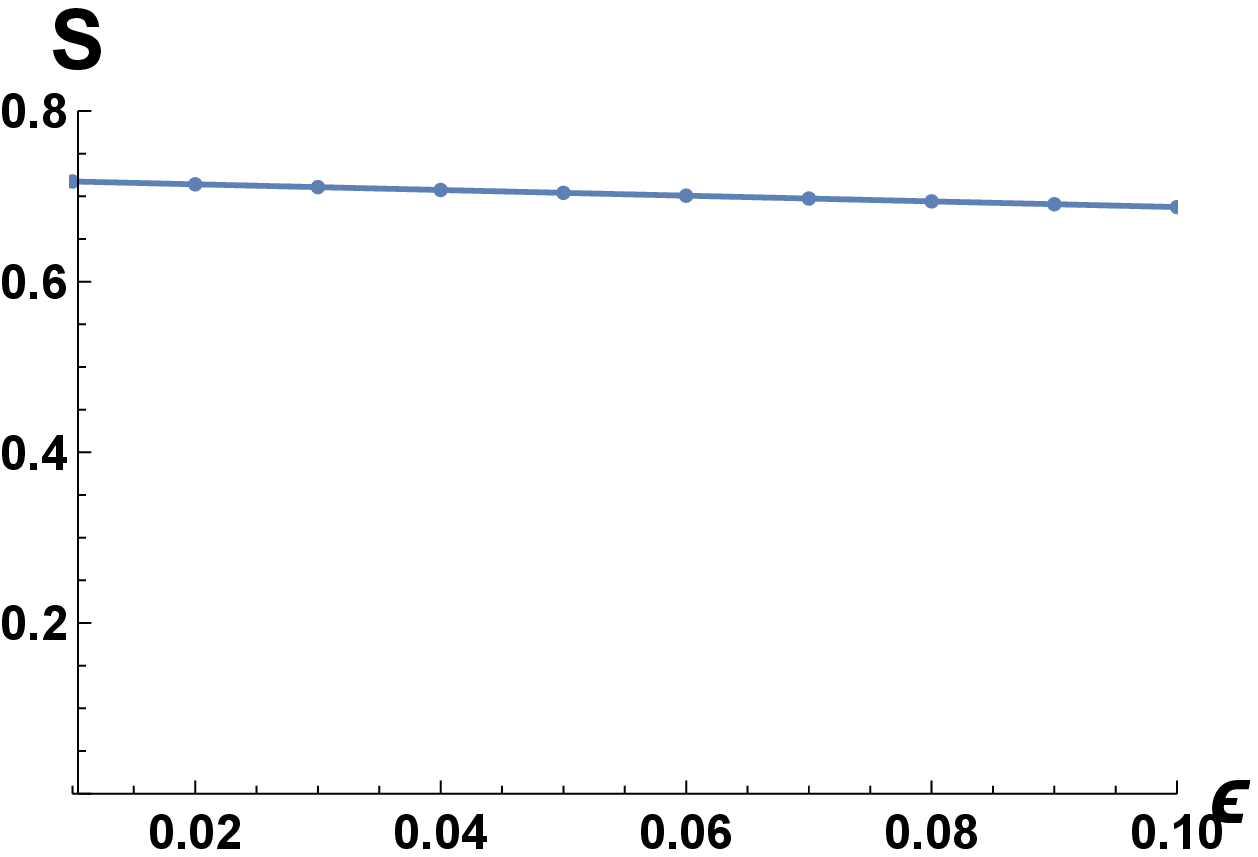}
	\caption{ 
The suprlus  $S$ versus
 $\epsilon$, 
  $f(x)=3x$, $L=500$.
{ The smooth line is given by numerics, the solid dots are the theoretical results by Eq. (17).}	}
\end{figure}

We have a distribution with 
double peaks.

For the linear fitness case $f(x)=kx $ we get
\begin{eqnarray}
\label{e18} 
&\rho(L)=\frac{\epsilon}{R+k+1+\epsilon},\nonumber\\
&R=\sqrt{k^2+1}-1-\epsilon,\nonumber\\
&s=\frac{ \sqrt{k^2+1}-1}{k},\nonumber\\
&1-p=\frac{\epsilon}{R+1+\epsilon}
\end{eqnarray}
Figures 4-5 illustrate the accuracy of our analytical results for the linear fitness case.

For the single peak fitness, with the peak fitness $J$ and fitness for other sequences,
we get for the mean fitness just
\begin{eqnarray}
\label{e19} 
R=J-1-\epsilon.
\end{eqnarray}

\subsection{The case of  resetting to the high fitness state with the smooth fitness function}

Consider the fitness choice   $f(x)=kx^2/2$, so {that} resetting is to 
the state with a maximal fitness.
We verified numerically that now we  have a single maximum for the distribution.
For the $f(x)=1.5 x^2$, we have at $\epsilon= 0$ case $R=2/3$.

Let us calculate the mean fitness of this phase.
First we simplify Eq. (6) at 
steady state for the $|L-l|\ll L$.
\begin{eqnarray}
\label{e20} 
&(f(m)-1-\epsilon)\rho(l)+\nonumber\\
&\mu(\frac{L-l+1}{L}\rho(l-1)
+\frac{l+1}{L}\rho(l+1)=0
\end{eqnarray}
We expand $ f(m)\approx f(-1)+ f'(-1)2n/L$, where $n=L-l$, then get an equation.
\begin{eqnarray}
\label{e21} 
&(f(-1)-R-1-\epsilon+  f'(-1)2n/L)\rho(L-n)+\nonumber\\
&\frac{(n+1)}{L}\rho(L-n-1)+\frac{(L-n+1)}{L}\rho(L-n+1) =0
\end{eqnarray}
In the bulk approximation, we obtain
\begin{eqnarray}
\label{e22} 
&\rho(L-n)=\frac{\rho (L-n+1)}{\Delta},\nonumber\\
&\Delta=-f(-1)+R+1+\epsilon
\end{eqnarray}
Thus Eq. (20) gives
\begin{eqnarray}
\label{e23} 
&(f(-1)-R-1-\epsilon+  f'(-1)2n/L)P(L-n)+\nonumber\\
&\frac{(n+1)}{L}\rho(L-n)
+\frac{(L-n+1)}{L}\rho(L-n+1).
\end{eqnarray}
Then
\begin{eqnarray}
\label{e24} 
&\rho(L-n)=\nonumber\\
&\rho (L-n+1)/(\Delta-\frac{(n+1)}{L\Delta}+\frac{(n-1)}{L\Delta}-kn)
\end{eqnarray}
where $ k=f'(-1)2/L $
We should look at an infinite product
\begin{eqnarray}
\label{e25} 
\rho(L-n)=\rho (L)\prod_m 1/(\Delta-\frac{(n+1)}{L\Delta}+\frac{(n-1)}{L\Delta}-kn), \nonumber\\
\end{eqnarray}
then calculate the sum
\begin{eqnarray}
\label{e26} 
\hat  P(R)=  \sum_{n=0} \rho (L-n).
\end{eqnarray}
Using the equation
\begin{eqnarray}
\label{e27} 
&F(a,b)=\sum_{n=0} \prod_{l=0}^n\frac{1}{a+bn}=\nonumber\\
&e^{1/b}\left (\frac{1}{b} \right)^{-\frac{a}{b}}
(\Gamma(\frac{a + b}{b} ) - \Gamma(\frac{a}{b} + 1, \frac{1}{b} ) ) + \frac{1}{a},
\end{eqnarray}
we then derive an {expression} 
for the mean fitness $R$.
\begin{eqnarray}
\label{e28} 
&F(a,b)=R,\nonumber\\
&a=\Delta-\frac{1}{\Delta L}-\frac{\Delta}{ L},\nonumber\\
&b=k-\frac{1}{\Delta L}+\frac{\Delta}{ L}
\end{eqnarray}

\begin{figure}[h!]
	\center
	\includegraphics[scale=0.5]{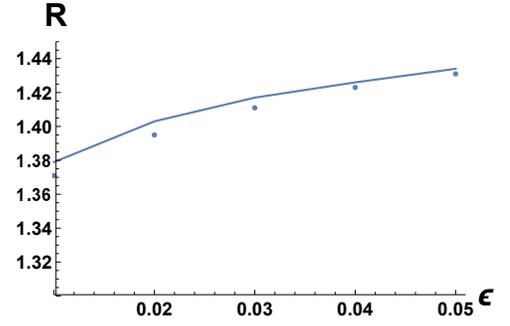}
	\caption{ 
The result of the mean fitness $R$ versus $\epsilon$ for the quadratic fitness
function $f(x) =3/2x^2, L= 1000$. 
The smooth line is given by our numeric, the solid dots are the result of our theory Eq. (28). 
For the $\epsilon = 0$ case, we have $R\approx 0.6666$. For $\epsilon=0.000001$ our theory gave $1.17$ versus $1.19$ by
numerics.
	}
\end{figure}

\begin{figure}[h!]
	\center
	\includegraphics[scale=0.5]{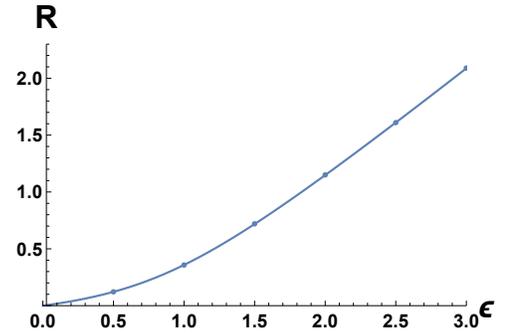}
	\caption{ 
The result of the mean fitness $R$ versus $J$ for the single peak fitness case, when the resetting is to the peak sequence, $J$ is the fitness of the mean sequence,
$\epsilon=0.2$.  The solid dots correspond o our numeric, the smooth line is given by Eq. (30).
	}
\end{figure}

We calculate $R$ using the last equation and gave the comparison with numerics in fig 6.
We verified that both cases on Figures 2,3 are given by the same Eq. (28).\\
{\subsection{The case of  resetting to the high fitness state with single peak fitness function}
Consider now the single peak fitness case, when the resetting sequence coincides with  the peak sequence.
We modify Eq. (6), considering the resetting to the 0-th Hamming class.  We have a fitness $J$ for the peak sequence, and 0 fitness for other sequences.
 Ignoring the $O(1/L)$ terms, we get
\begin{eqnarray}
\label{e29} 
&\rho(0)R=(J-1)\rho(0)+\epsilon(1-\rho(0)),\nonumber\\
&R=J \rho(0)
\end{eqnarray}
Then we derive the following equation for the mean fitness in the steady state
\begin{eqnarray}
\label{e30} 
R=\frac{J-1+\epsilon+\sqrt{(J-1+\epsilon)^2+4 J\epsilon}}{2J}
\end{eqnarray}
}
\section{Conclusion}
Random walk models with stochastic resetting are {on}
the focus of the modern statistical physics,
and have numerous applications.
{ While there have been some resetting related  results in population genetics, till now the quasispecies models have not been investigated in case of stochastic resetting. }
We formulated the resetting version of the Crow-Kimura model  Eq. (6), 
then investigated the model both numerically and analytically.
It is {astonishing}
funny 
that 
stochastic resetting arose {for the} first time just in 
population genetics \cite{fl79}, so our work the returns back the research process to the origins.
We identify two statistical physics phases in our model.
The first one { given by Eqs. (10),(15)-(17)}  is not too hard mathematical problem, this phase  in the case when the resetting sequence is an ordinary sequence
with a non-highest fitness. Our analytics is well confirmed by numeric, see fiigures (4)-(5).
{
When we choose {as} the resetting sequence the sequence with the fitness peak, 
the situation is becoming highly non-trivial, { the solution is given byhypergeometric function Eq. (28) }. We met a {highly}
non-perturbative phenomenon.
Even the $\sim10^{-6}$ resetting probability brings to $100\%$ change of the mean fitness.
{ Eqs. (6),(7) illustrate the accuracy of our analytical results.}

It will be interesting to look the equivalents of our findings in other stochastic resetting models, as 
the driving forces in random walks resemble our fitness in case of evolution models.
Here we looked at the most direct mapping of resetting mechanism from the random walks to evolution.
An alternative approach should be {to look at}
the evolution models with reservoirs - there is a two-habitat model \cite{ev,sa18} with some transitions between
them, and 
second, reservoir habitat {where}
the mutations are suppressed by strong selection forces.

This work was supported by SCS of Armenia, grants  No. 20TTAT-QTa003 and 21T-1C037.

\renewcommand{\theequation}{A.\arabic{equation}}
\setcounter{equation}{0}

\end{document}